\documentclass[11pt,a4paper]{iopart}
\usepackage{iopams}
\usepackage{setstack,cite}
\usepackage{amssymb,graphicx}
\usepackage{tabularx}
\usepackage{float}

\newcommand{\bea}{\begin{eqnarray}}
\newcommand{\eea}{\end{eqnarray}}
\newcommand{\bes}{\numparts}
\newcommand{\ees}{\endnumparts}

\begin{document}
\bibliographystyle{iopart-num}
\title[A systematic construction of $\cal PT$-symmetric and non-$\cal PT$-symmetric complex potentials ...]{A systematic construction of parity-time ($\cal PT$)-symmetric and non-$\cal PT$-symmetric complex potentials from the solutions of various real nonlinear evolution equations}
\author{K. Tamilselvan$^1$, T. Kanna$^1$ and Avinash Khare$^2$}
\address{$^1$Post Graduate and Research Department of Physics, Bishop Heber College, Tiruchirappalli--620 017, Tamil Nadu, India}
\address{$^2$Physics Department, Savitribai Phule Pune University \\
Pune, India 411007}
\eads{kanna\_phy@bhc.edu.in (corresponding author)}

\begin{abstract}
We systematically construct a distinct class of complex potentials including  parity-time ($\cal PT$) symmetric potentials for the stationary Schr\"odinger equation by using the soliton and periodic solutions of the  four integrable real nonlinear evolution equations (NLEEs) namely the sine-Gordon (sG) equation, the modified Korteweg-de Vries (mKdV) equation, combined mKdV-sG equation and the Gardner equation. These potentials comprise of kink, breather, bion, elliptic bion, periodic  and soliton potentials which are explicitly constructed from the various respective solutions of the NLEEs. We demonstrate the relevance between the identified complex potentials and the potential of the graphene model from an application point of view.
\end{abstract}
\maketitle
\section{Introduction}
Over the last 15 years, the study of parity-time ($\cal PT$) symmetry has received wide attention  in different branches of physics as they offered strong physical and mathematical insights. Initially, it  emerged in quantum mechanics, where it has been shown that even if a Hamiltonian is not hermitian but admits $\cal PT$-invariance, then the energy eigenvalues are still real in case the $\cal PT$ symmetry is not broken spontaneously. This phenomenon was first observed in the pioneering work of Bender and Boettcher \cite{bender}. The fascinating properties of $\cal PT$-symmetric systems have paved the way for numerous developments in diverse areas of physics namely, quantum field theory, optics, photonics and Bose-Einstein condensates. Particularly, the concept of $\cal PT$ symmetry has inspired  tremendous advances in optics because it has provided  a multifaceted platform. In the realm of linear optics, theoretically complex $\cal PT$-symmetric potential has been studied in \cite{Gnanainy} followed by several experimental works on different optical systems, to name a few, coupled waveguide system \cite{ruter}, synthetic materials \cite{peng} and so on.

The next natural step is to explore various phenomena in nonlinear optical systems with $\cal PT$-symmetry. In nonlinear optics due to the interplay between the $\cal PT$ symmetric potentials and dispersion/diffraction as well as nonlinear effects, different types of localized structures (these
structures are of relevance to the $\cal PT$ symmetric property) arise, such as solitons \cite{mussli, abdulla, he, moreira}, periodic waves \cite{truong}, gap solitons \cite{moreira1} and vortices \cite{achilleos}. Particulary, the existence of different nonlinear localized modes has been studied analytically as well as numerically in the nonlinear Schr\"odinger (NLS) equation with complex $\cal PT$ symmetric periodic potential \cite{musslimani0}, Scarf-II potential\cite{Li}, Gaussian potential \cite{Hu}, Bessel potential \cite{Hu1}, Rosen-Morse Potential \cite{Bikashkali}, harmonic potential\cite{zezyulin0}.  In addition to these, nonlinear modes have been studied for other complex $\cal PT$- symmetric potentials bearing nonlinear optical systems like competing nonlinearity \cite{avinash1}, saturable nonlinearity\cite{honga}, logarithmically saturable nonlinearity \cite{zhank}. On the other hand the complex potentials are used in BEC also to explore nonlinear localized modes \cite{achileos0}.

These diversified studies clearly emphasise the need for the development of analytical procedure to construct many more $\cal PT$-symmetric potentials. In a classic work \cite{wadati},  Wadati has developed an interesting method to construct $\cal PT$-symmetric potentials of the form
\bea\label{f0}
V(x)&=&-q^{2}-i q_{x},
\eea
where the real function $q(x)$ is referred as potential base \cite{wadati}. Recently, Wadati potentials have stated to receive renewed attention. Especially, the symmetry breaking of solitons has been observed in a class of non-$\cal PT$ Wadati like potentials beyond certain threshold value \cite{yangj}. In another interesting work \cite{tsoy}, nonlinear modes have been realized for the asymmetric waveguide profiles with Wadati potential. Stability analysis was also performed numerically for the Wadati like non-$\cal PT$ symmetric potentials \cite{yangj1}. More recently, the exact solutions of the Gross-Pitaevskii equation with Wadati like potential are constructed in \cite{zezyulin}.

Inspired by these recent advances, we obtain here a wide class of complex potentials including  $\cal PT$ symmetric potentials for the linear Schr\"odinger equation with the aid of soliton and periodic solutions of different real integrable nonlinear evolution equations (NLEEs) namely the sine-Gordon (sG) equation, the modified Korteweg de Vries (mKdV) equation, combined mKdV-sG equation and the Gardner equation. All these equations are of considerable physical interest. These physically interesting NLEEs describe a plethora of phenomena in nonlinear science. For instance the sG system arises in a wide variety of physical systems, such as, long Josephson junction placed in an alternating electromagnetic field, elementary excitations of weakly pinned Fr\"ohlich charge-density-wave condensates at low temperatures, DNA double helix \cite{mineev} etc. The mKdV equation is useful in the study of the dust ion acoustic solitary waves in unmagnetized dusty plasma, Alfve\'n solitons in relativistic electron-positron plasma, the propagation of solitary waves in Schottky barrier transmission lines, in the models of traffic congestion \cite{lonngren} etc. Further, the mKdV-sG equation can be employed to describe physical situations such as nonlinear wave propagation in an infinite one dimensional anharmonic lattice and the propagation of ultrashort optical pulses solitons (or few cycle pulses solitons) in a Kerr medium \cite{leblond1}. Finally, the Gardner equation is useful in the study of propagation of ion-acoustic waves in plasmas and internal waves in a stratified ocean\cite{ruderman}. As mentioned above, optical systems can act as fertile ground for $\cal PT$ -symmetry properties. One of the main advantages of the systems that are presently under consideration in this paper is that except the Gardner equation, all the remaining equations arise in the context of nonlinear optics. Particularly, the propagation of few cycle pulses in Kerr media can be well described by using the mKdV, sG and mKdV-sG equations in the non-slowly varying envelope
approximation \cite{melnikov}.

In this paper we adapt the methodology developed by Wadati \cite{wadati} to construct $\cal PT$ symmetric potentials. This method exploits the connection between time independent Schr\"odinger equation and the Zakharov-Shabat (Z-S) spectral problem to construct $\cal PT$-symmetric potentials using the several known periodic as well as hyperbolic soliton solutions of the above mentioned integrables NLEEs. In his pioneering work, Wadati has shown that the spatial evolution equations of the Z-S eigenvalue problem
\bes\label{f8b}\bea
v_{1x}+ i \zeta v_{1}&=&q v_{2}, \\\label{f8c}
v_{2x}-i \zeta v_{2}&=&- q v_{1}
\eea\ees
corresponding to the mKdV equation
\bea\label{f8}
q_{t}+6 q^{2}q_{x}+q_{xxx}=0,
\eea
can be transformed into the linear Schr\"odinger type eigenvalue problem featuring complex potential of the form (\ref{f0}) with eigenvalues being the square of the spectral parameter $\zeta^{2}$ of the Z-S problem. This complex potential is then explicitly constructed with the aid of the exact solutions of mKdV equation.  The main point is that for those solutions which are derived from the IST method, one can immediately figure out if the energy eigenvalues of the stationary Schr\"odinger-like equation corresponding to these complex potentials are real or appear in complex conjugate pairs. This is an interesting result in the context of the complex $\cal PT$-invariant potentials as in that case one can then say if the corresponding eigenfunctions of the Schr\"odinger equation are $\cal PT$ invariant or not, i.e. if the $\cal PT$-symmetry is spontaneously broken or not. Here we wish to note that in a recent work \cite{barashenkov},  Barashenkov I V \textit{et al}., have developed a procedure to construct $\cal PT$-symmetric Wadati type potentials for the Gross- Pitaevskii equation (nonlinear Schr\"odinger equation). In Ref.\cite{barashenkov}, the stationary nonlinear Schr\"odinger equation with unknown potential of Wadati type is systematically solved to explicitly construct the exact form of the Wadati potential and the nonlinear mode exhibiting such potentials is also found. In the present work, we have constructed general complex potentials of the linear Schr\"odinger equation with the aid of various known solutions of the integrable NLEEs like sG equation, mKdV equation, combined mKdV-sG equation and Gardner equation as explained above.

For the majority of the cases considered in this paper, one obtains $\cal PT$ invariant complex potentials but in certain cases one also obtains complex potentials which are not $\cal PT$ invariant. Especially, we consider the IST solutions for constructing complex potentials corresponding to the sG and the mKdV-sG systems. On the otherhand, Wadati has already considered \cite{wadati} the potential based on the IST method solutions of the mKdV and the Gardner systems and constructed the corresponding complex potential as mentioned before. Hence, for completeness, we only focus our attention on the interesting solutions of the mKdV and the Gardner equation which are not obtained by the IST method. Finally, we also point out the possible relevance of some of the complex $\cal PT$-invariant solutions of the sG and the mKdV-sG systems in the context of the graphene system which is described by the Dirac equation. It is worth pointing out here  that in the context of the same graphene system, the relevance of the complex $\cal PT$-symmetric potentials following from the solutions of the mKdV and the Gardner equations has been
shown before \cite{ho}.

This paper is organized as follows: In sec. 2, we construct $\cal PT$ symmetric potentials, both periodic and hyperbolic types, from the solution of sG equation. We show that in the special case of  the breather solutions we obtain potentials which do not admit $\cal PT$ symmetry. We also show here
the possible relevance of the obtained potentials from the sG equation in the context of the graphene model. In sec. 3, we apply the systematic method discussed in \cite{wadati} to the mKdV equation and construct $\cal PT$ symmetric periodic as well as hyperbolic potentials. We then consider the combined mKdV-sG and
Gardner equations in secs. 4 and 5 respectively and in both the cases we obtain complex $\cal PT$ invariant potentials. Our results are summarized in sec. 6.\\

\section{Complex potentials from the solution of sine-Gordon equation}

To start with, let us consider the ubiquitous sine-Gordon equation
\cite{rajaraman,cuevas}
\bea\label{f1}
u_{tt}-u_{xx}+\sin u=0\,.
\eea
Here, the field $u(x,t)$ is real and the subscripts $x$ and $t$ denote partial
derivatives with respect to space and time respectively. The field of the sG
equation satisfies vanishing boundary condition, i.e., $u(x,t)\rightarrow 0$
at $x\rightarrow \pm \infty$. Eq. (\ref{f1}) results from the compatibility
condition of the famous Zakharov-Shabat (Z-S) $(2\times 2)$ problem with the
following spatial and temporal evolution equations:
\bes\bea
v_{1x}+ i \zeta v_{1}&=&-q v_{2},\,\label{f2a}\\\label{f2b}
v_{2x}-i \zeta v_{2}&=&q v_{1}\,
\eea
and
\bea\label{f2c}
v_{1t}=\left(\frac{i}{4\zeta}\right)\left(v_{1} \mbox{cos(u)}
+ v_{2}\mbox{sin(u)}\right),\,\\
v_{2t}=\left(\frac{i}{4\zeta}\right)\left(v_{1} \mbox{sin(u)} -
v_{2}\mbox{cos(u)}\right),\,
\eea\label{f2}\ees
where, $q=\frac{u_{x}}{2}$ and $\zeta$ is the spectral parameter which is
in general complex. One can arrive at the following one dimensional stationary
Schr\"odinger equation involving sine-Gordon potential $V^{(sG)}$ by
differentiating Eqs. (\ref{f2a}) and (\ref{f2b}) with respect to $x$ and by
introducing new eigenfunctions $\phi_{1}=v_{2}-i v_{1}$ and
$\phi_{2}=v_{2}+i v_{1}$:
\bes\bea
-\frac{d^2\phi_{1}}{d x^{2}}+V^{(sG)} \phi_{1}& =&\zeta^{2}\phi_{1},\,
\label{f4}\\ \label{f4a}
-\frac{d^2\phi_{2}}{d x^{2}}+V^{*(sG)} \phi_{2}& =&\zeta^{2}\phi_{2},\,
\eea\ees
where $*$ denotes the complex conjugation.
The exact expressions for $V^{(sG)}$ is given by,
\bea
V^{(sG)}=-\frac{(u_{x})^{2}}{4}-i \frac{u_{xx}}{2} \equiv V_{R}(x) + i V_{I}(x)\,.\label{f4b}
%V^{*(sG)}=-\frac{(u_{x})^{2}}{4}+i \frac{u_{xx}}{2} = V_{R}(x) -i V_{I}(x)\,.
\eea
Thus the Hamiltonian consists of the complex potential that is furnished by
the exact solution $u(x,t)$ of the sG equation. Though $u(x,t)$ depends on
$t$, the energy spectrum of the Z-S problem (i.e. $\zeta^2$)  is independent
of $t$, which implies that $t$ can be viewed as a deformation parameter.
Correspondingly the various solutions  $u(x, t)$ of the sG Eq. (\ref{f1})
generate deformable potentials $V(x,t)$ with real spectra if $\zeta$ is
either real or purely imaginary. Hence the initial time and the solution
parameters can be chosen appropriately such that solution is independent of time
during the construction of potentials \cite{konotop}. It should be noted that
the real and imaginary parts of the complex potential associated with the sG equation are
obtained from the real function $u(x,t)$. Thus it is the real solution
$u(x,t)$ of the sG Eq. (\ref{f1}) which determines the complex potential
$V^{(sG)}(x,t)$ of the sG equation \cite{wadati}.

We remark that the real energy eigenvalues of the sG equation imply unbroken
$\cal PT$ -symmetry whereas the complex conjugate pairs of energy eigenvalues
of the sG equation lead to the spontaneous breaking of the $\cal PT$ -symmetry.

\subsection{Relevance of the sG in the context of Dirac equation for the
Graphene model}

At this stage it is worth pointing out one possible application of the complex
potentials obtained from the various solutions of the sG equation. In this
regard, we wish to point out that very recently in Ref \cite{ho} a close connection has been
established between the Dirac equation for the graphene model and the mKdV and
the Gardner equations. Such a connection was established by showing
that the solutions of these NLEEs can actually act as electrostatic
potentials of the Dirac equation for the charge carriers in graphene at zero
energy state. Along those lines, in order to realize the existence of similar
connection between the graphene model and the solution of the sG potential, we first
consider the following equation describing the motion of electrons in graphene
in the presence of an electrostatic field \cite{Del}
\bea
\left[v_{F}(\sigma_{x} p_{x}+\sigma_{y}p_{y})\right]\psi+U(x,y)\psi=E\psi.
\label{f5b}
\eea
Here $p_{x}=-i\frac{\partial}{\partial x}$, $p_{y}
=-i\frac{\partial}{\partial y}$, $v_{F}$, $U(x,y)$ and $E$ are the momentum
operators in the $x$ and $y$ directions, the Fermi velocity, the potential
energy and energy eigenvalue respectively. The wavefunction is chosen to be of
the following form
\bea
\psi(x)=\Bigg(
  \begin{array}{c}
    \psi_{A} \\
    \psi_{B} \\
  \end{array}
\Bigg)e^{ik_{y}y},
\label{f5c}
\eea
where $k_{y}$ is the wave vector of the $y$-component and the potential $U$ is
assumed to depend only upon the $x$-coordinate. Substituting Eq. (\ref{f5c})
into Eq. (\ref{f5b}) and introducing the new eigenfunctions
$\psi_{1,2}=(\psi_{A}\pm\psi_{B})$, we obtain
\bes\bea
-\frac{d^2\psi_{1}}{d x^{2}}-\left((V (x)-\epsilon)^{2}+i \frac{dV}{dx}\right)\psi_{1}=-k_{y}^{2} \psi_{1},\label{f5d}\\ \label{f5e}
-\frac{d^2\psi_{2}}{d x^{2}}-\left((V (x)-\epsilon)^{2}-i \frac{dV}{dx}\right)\psi_{2}=-k_{y}^{2} \psi_{2},
\eea\ees
where $V(x)=\frac{U(x)}{v_{F}}$ and $\epsilon=\frac{E}{v_{F}}$. To bring out the connection between  Eqs.~(10) and Eqs.~(6), for zero energy states of graphene ($\epsilon=0$), we make the replacement $\psi_{1,2}\leftrightarrow \phi_{1,2} $ and $-k^2_{y}\leftrightarrow \zeta ^2$. Then we notice that the resulting equations become identical for the choice $V(x)=\frac{u_{x}}{2}$. This gives the connection between the potential of the graphene model and the solution $u$ of the sG equation. In its original variables, the potential reads as
\bea\label{f5g}
U(x)=\frac{v_{F} u_{x}}{2}.
\eea
Note that this graphene potential is distinctly different from the previously
reported potentials obtained from the mKdV and the Gardner system in \cite{ho}.
This clearly demonstrates that the various solutions of the sG equation can
furnish distinct exactly solvable electrostatic potentials for the graphene
system corresponding to zero energy states. Going one step further than in
\cite{ho}, we find that for arbitrary energy states (with particular value for
$\epsilon$) the connection between Eqs. (\ref{f4}) and (\ref{f4a}) and Eqs. (\ref{f5d}) and (\ref{f5e}) holds
good for $ V(x)=\frac{u_{x}}{2}+\epsilon$.

\subsection{Construction of complex potentials}
Now, we return to our main task and proceed to construct the complex potentials
 by using the exact solutions $u(x,t_{0})$, (where $t_{0}$ is the initial time) of the sG equation. The integrable sG equation admits the following N-kink soliton solution which is obtained by the IST method \cite{Ablowitz}
\bea
u(x,t)&=&\cos^{-1}\left[1+2\left(\frac{\partial^{2}}{\partial t^{2}}-\frac{\partial^{2}}{\partial x^{2}}\right)\ln f(x,t)\right],\,\label{f6}
\eea
where $f(x,t)=\det(M(x,t))$ in which the elements $M_{ij}$ of the $N\times N$ square matrix $M$ can be
expressed as $\frac{\left(a_{i}a_{j}\right)^{\frac{1}{2}}}{a_{i}+a_{j}}
\left[e^{\theta_{i}}+(-1)^{i+j}e^{-\theta_{j}})\right]$ in which $\theta_{i}$
= $\frac{1}{2}\left(a_{i}+\frac{1}{a_{i}}\right)x+\delta_{i}(t)$ and
$\delta_{i}(t)$ can be expressed as  $\frac{1}{2}\left(a_{i}
-\frac{1}{a_{i}}\right)t+\rho_{i}~~(i,j=1,2,...,N)$ with arbitrary constants
$a_{i}$ and $\rho_{i}$. Remarkably, it is well known that for the $N$-soliton
solution the parameter $\zeta$ is pure imaginary and hence the energy
$\zeta^2$ of the Schr\"odinger-like Eq. (\ref{f4}) is real. In the case of one and two soliton solutions we now explicitly show that the corresponding complex potential $V^{(SG)}$ is $\cal PT$ invariant
thereby demonstrating that in these cases the $\cal PT$ symmetry is not
spontaneously broken. We conjecture that the same is also true in the
case of $N$ soliton solutions of the sG Eq. (\ref{f1}). In addition to that, one can use the recently reported degenerate multi-soliton and multi-breather solutions \cite{cen}, to construct the complex potentials.

\subsubsection{\bf One-kink (or Scarf-II) potential for $N=1$}

We now write down the exact one-kink solution of the sG equation resulting
from Eq. (\ref{f6}) for $N=1$
\bea\label{f6a}
u(x,t)=4\tan^{-1}\left(e^{\theta_{1}}\right)\,,
\eea
where, $\theta_{1}=\frac{1}{2}\left(a_{1}+\frac{1}{a_{1}}\right)x
+\delta_{1}(t)$, $ \delta_{1}(t)=\frac{1}{2}\left(a_{1}
-\frac{1}{a_{1}}\right)t+\rho_{1}$. By choosing the initial time $t_{0}$ such that $\delta_{1}(t_{0})=\frac{1}{2}\left(a_{1} -\frac{1}{a_{1}}\right)t_{0}+\rho_{1}=0$, we obtain the same solution (\ref{f6a}) but with $\theta_{1}=\frac{1}{2}\left(a_{1}+\frac{1}{a_{1}}\right)x$. Then from (\ref{f4b}) we construct the one-kink
potential as
\bea\label{f6abc}
V^{(sG)}(x)=-\frac{1}{4}\left(a_{1}+\frac{1}{a_{1}}\right)^{2}\Big[\mbox{sech}^{2}(\theta_{1})+i\tanh(\theta_{1})
~\mbox{sech}(\theta_{1})\Big].
\eea
This type of potential is referred as ($\cal PT$-symmetric) Scarf-II potential. It has
other names too such as hyperbolic Scarf potential or Gendenshtein potential
\cite{Gendenshtein}. The Scarf-II potential is characterized by the same real component as that of complex Rosen-Morse potential but with different imaginary part.  Interestingly, defocusing Kerr media featuring Scarf-II potential can support bright-soliton\cite{shi}. Possibility of gray soliton in defocusing media with Scarf-II potential has also been analysed numerically in \cite{Li}. This Scarf-II potential is clearly $PT$-invariant. Further, the $\cal PT$-symmetry is unbroken spontaneously since this potential (which admits one bound state) has real energy eigenvalue $E^{(sG)}=-a_{1}^{2}$, since from the IST method it is well known that $\zeta=i a_{1}$.

It may be noted here that in spite of the alteration of the parameter $a_{1}$,
the shape of the Scarf-II potential remains unchanged. Further, the real and
the imaginary parts of the Scarf-II potential vanish at $x = \pm\infty$.

\begin{figure}[H]
\centering\includegraphics[width=0.4\linewidth]{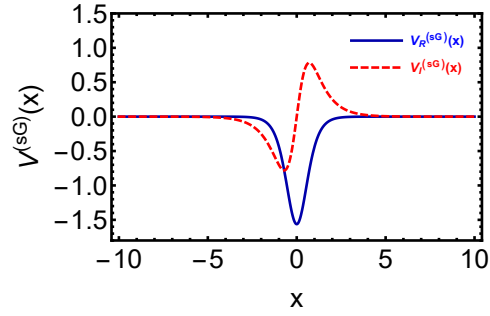}~~~~~
\caption{Real part (solid blue line) and imaginary part (red-dashed line) of the one-kink potential (\ref{f6abc}) for $a_{1}=-0.5$.}
\end{figure}
To illustrate the nature of the potential, we display, the real part (solid
blue line) and imaginary part (red-dashed line) of the Scarf-II potential in
Fig.~1.

\subsubsection{\bf Two-kink potential for $N=2$}
Next we shift our attention to the $N =2$ case, i.e. two-kink soliton solution of
the sG equation obtained from Eq. (\ref{f6})
\bea\label{f6b}
u(x,t)=4 \tan^{-1}\left[a\sinh\left(b\right)\mbox{sech} \left(c\right)\right],
\eea
where, $a=\left(\frac{a_{1}-a_{2}}{a_{1}+a_{2}}\right)$; $b=\frac{1}{2}\left(\theta_{1}+\theta_{2}\right)$ and
$c=\frac{1}{2}\left(\theta_{1}-\theta_{2}\right)$. Here $\theta_{1,2}$ is
expressed as $\frac{1}{2}\left(a_{1,2}+\frac{1}{a_{1,2}}\right)x
+\delta_{1,2}(t)$, $\delta_{1,2}(t)=\frac{1}{2}\left(a_{1,2}
-\frac{1}{a_{1,2}}\right)t+\rho_{1,2}$. The time dependence in (\ref{f6b}) can
be eliminated by choosing the initial time $t=t_{0}$, as
$\delta_{1,2}(t_{0})=\frac{1}{2}\left(a_{1,2}-\frac{1}{a_{1,2}}\right)t_{0}
+\rho_{1,2}=0$.  The resulting solution bears the same form as in
(\ref{f6b}), while $\theta_{1,2}=\frac{1}{2}\left(a_{1,2}
+\frac{1}{a_{1,2}}\right)x$ and ultimately, the parameters $b$ and $c$ are
time independent. Then, we construct two-kink potential from (\ref{f4b}) as
\bea\label{f6ba}
V^{(sG)}(x)=\frac{1}{4}\Bigg[\Bigg(4a\frac{\big(b~\mbox{sech}(c)\cosh(b)-c~\sinh(b)\mbox{sech}(c)\tanh(c)\big)}{1+\big(a\sinh(b)\mbox{sech}(c)\big)^{2}}\Bigg)^{2}\nonumber\\
-i\Bigg(2\frac{\mbox{d}}{\mbox{dx}}\Bigg(4a\frac{\big(b~\mbox{sech}(c)\cosh(b)-
c~\sinh(b)\mbox{sech}(c)\tanh(c)\big)}{1+\big(a\sinh(b)
\mbox{sech}(c)\big)^{2}}\Bigg)\Bigg)\Bigg].
\eea
Note that this two-kink potential is also $\cal PT$ -invariant
and the $\cal PT$ -symmetry is unbroken since from the IST method it is well known that
this potential admits two bound states with real energy
eigenvalues $E^{(sG)}_{1}=-a^{2}_{1}$ and $E^{(sG)}_{2}=-a^{2}_{2}$.

It is amusing to note that unlike the one-kink potential, the shape of the
two-kink potential can be altered by changing the values
of the soliton parameters $a_{1}$ and $a_{2}$.
\begin{figure}[H]
\centering\includegraphics[width=0.4\linewidth]{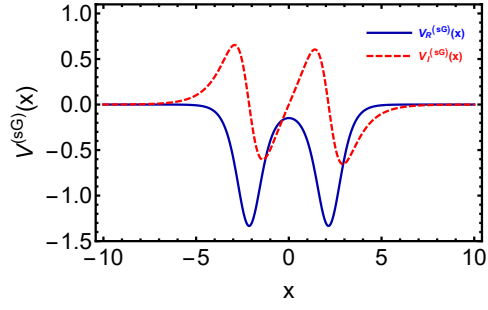}
\caption{Real part (solid blue line) and imaginary part (red-dashed line) of the two-kink potential (\ref{f6ba}) for $a_{1}= 0.5$ and $a_{2} = 0.7$.}
\end{figure}
In order to analyse the nature of the two-kink potential, we have plotted the
real part (continuous blue line) and the imaginary part (red-dashed line) of
the two-kink potential in Fig.~2.

\subsubsection{\bf  Breather type potential}

Next, we move on to construct potential following from the celebrated breather
solution of the sG Eq. (\ref{f1}). In this connection, we start with the
 breather solution of the sG equation \cite{Ablowitz} obtained by the IST method:
\bea\label{f6c}
u(x,t)=4 \tan^{-1}\left(a_{0}~\mbox{sech}\left(a_{3} x+\delta_{3}\right)~\cos \left(a_{4} x+\delta_{4}\right)\right),
\eea
where $a_{0}=\left(\frac{\eta}{\xi}\right)$, $a_{3}=\left(\frac{\eta~\nu}{2}\right)$ and $a_{4}=\left(\frac{\xi(\nu-4)}{2}\right)$, in which  $\eta$, $\xi$ and $\nu$ are real parameters, $\delta_{3}$ and $\delta_{4}$ are expressed as $-(4-\nu)t- a_{3} x_{0}\nu$ and $\left(\frac{\xi}{2}\right)\nu (t-\tau_{0})$ respectively. Here, $x_{0}$ and $\tau_{0}$ are arbitrary constants. By considering the initial time $t_{0}$ such that $\delta_{3}(t_{0})= -(4-\nu)t_{0}- a_{3} x_{0}\nu=0$ and $\delta_{4}(t_{0})=\left(\frac{\xi}{2}\right)\nu (t_{0}-\tau_{0})=0$,  the solution (\ref{f6c}) becomes
\bea\label{f6d}
u(x)=4 \tan^{-1}\left(a_{0}~\mbox{sech}\left(a_{3}\right)\cos \left(a_{4}x\right)\right).
\eea
The breather solution (\ref{f6d}) is obtained by the IST method which
involves complex conjugate pairs of eigenvalues. The breather potential
resulting from solution (\ref{f6d}) is given below,
\bea\label{f6e}
V^{(sG)}(x)= -\frac{(4~a_{0}^2~\mbox{sech}^{2}(a_{3} x)~(a_{4}~\sin(a_{4} x) + a_{3}~\cos(a_{4} x)\tanh(a_{3} x))^{2})}{(1 +
   a^2 \cos^{2}(a_{4} x) \mbox{sech}^{2}(a_{3} x))^{2}}+
\nonumber\\
+2i\frac{\mbox{d}}{\mbox{dx}}\Bigg(\frac{(a_{0}~\mbox{sech}(a_{3} x)~(a_{4}~\sin(a_{4} x) + a_{3}~\cos(a_{4} x)\tanh(a_{3} x))}{(1 +
   a^2 \cos^{2}(a_{4} x) \mbox{sech}^{2}(a_{3} x)}\Bigg)\Bigg],\qquad\qquad
\eea
The above complex breather-type potential does not have the
$\cal PT$ -symmetry.
This non-$\cal PT$ -symmetric potential admits robust shape for all real
values of  $\xi$ and $\eta$. One can easily notice that the real part and
imaginary parts of this breather potential asymptotically vanish.
It is amusing to note that while the soliton solutions lead to a potential
with $\cal PT$-symmetry, the breather solution fails to result in such
$\cal PT$ -symmetric potential.
This is of course perfectly understandable as the breather solution
(\ref{f6c}), is obtained by considering the eigenvalues as complex conjugate
pairs which results in complex spectral parameters $\zeta_{1}=\xi+ i \eta$
and $\zeta_{2}=\xi- i \eta$. The corresponding eigenvalues are thus
$E_{1}=\xi^{2}-\eta^{2} +2 i \xi \eta$ and
$E_{2}=\xi^{2}-\eta^{2}-2 i \xi \eta$.
\begin{figure}[H]
\centering\includegraphics[width=0.4\linewidth]{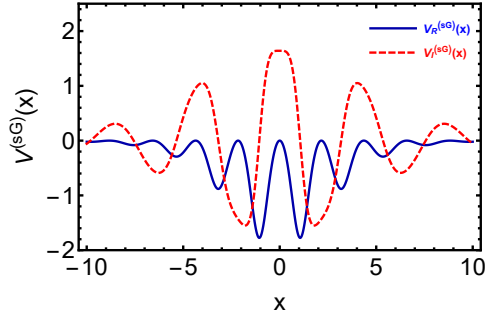}
\caption{Real part (solid blue line) and imaginary part (red-dashed line) of the breather potential (\ref{f6e}) for $\eta=0.5$, $\xi=1$, $\nu = 1.2$.}
\end{figure}
In Fig.~3, we depict the real part (solid blue line) and imaginary part
(red-dashed line) of breather type potential respectively.
Notice that both the real and the imaginary parts are symmetric function.

\subsubsection{ \bf Periodic two kink potential }

Now, we focus our attention to construct periodic two kink potential from
the periodic two kink solution of the sG equation. The periodic two kink
solution of the sG equation (\ref{f1}) can be written as \cite{Fu}
\bea\label{f7a}
u(x,t) = 4\tan^{-1}[\mbox{sd}(a x + \delta_{5}(t), k)~
\mbox{cn}(c x+\delta_{6}(t), m)]\,.
\eea
The constraint conditions are expressed as $a=m~c$;
$b=-d=\Big(\frac{m^2}{2m^2k^2-3m^2+1}\Big)$ and
$m=\pm\frac{1}{\sqrt{1+k^{2}-k^{4}}}$. Here $\delta_{5}(t)$ and $\delta_{6}(t)$
 are given by $b t+s_{0}$ and $d t+r_{0}$ respectively, in which $s_{0}$ and
$r_{0}$ are arbitrary constants. Here \mbox{cn}(cx,k), and
$\mbox{sd}(bx,m)\Big(\equiv \frac{\mbox{sn}(bx,m)}{\mbox{dn}(bx,m)}\Big)$ are Jacobi
elliptic functions with modulii \mbox{k} and \mbox{m}, respectively. These
elliptic functions $\mbox{sn}$, $\mbox{cn}$ and $\mbox{dn}$  are doubly periodic functions
having periods $4K$, $4K$ and $2K$ respectively. One can easily obtain the
following solution from solution (\ref{f7a}) by considering the initial time
$t_{0}$ such that $\delta_{5}(t_{0})=b t_{0}+s_{0}=0$ and $\delta_{6}(t_{0})
=d t_{0}+r_{0}=0$
\bea\label{f7aa}
u(x) = 4\tan^{-1}[\mbox{sd}(a x , k)~\mbox{cn}(c x, m)]\,.
\eea
Using solution (\ref{f7aa}), we obtain the following complex potential
\bes\bea
V^{(SG)}(x)=V^{(sG)}_{R}(x)+i V^{(sG)}_{I}(x)\,,
\label{f7b}\eea
where the real and imaginary parts of the periodic potential from the periodic solution of the sG equation are given respectively by
\bea\label{f7c}
V^{(sG)}_{R}(x)=-4\Big(\frac{a~\zeta_{1}~\mbox{nd}(ax,k)-c~\zeta_{2} ~\mbox{sn}(cx,m)}{(1+\mbox{cn}^{2}(cx,m)~\mbox{sd}^{2}(ax,k))}\Big)^{2}
\eea
and
\bea\label{f7d}
V^{(sG)}_{I}(x)=-2\frac{\mbox{d}}{\mbox{dx}}\Big(\frac{a~\zeta_{1}~\mbox{nd}(ax,k)-c~ \zeta_{2}~\mbox{sn}(cx,m)}{(1+\mbox{cn}^{2}(cx,m)~\mbox{sd}^{2}(ax,k))}\Big).
\eea\ees
Here $\zeta_{1}=\mbox{cd}(ax,k)~\mbox{cn}(cx,m)$, $\zeta_{2}=\mbox{dn}(cx,m)~\mbox{sd}(ax,k)$ and also $\mbox{cd}(bx,k)\Big(\equiv \frac{\mbox{cn}(bx,k)}{\mbox{dn}(bx,k)}\Big)$ and
$\mbox{nd}(bx,k)\Big(\equiv \frac{1}{\mbox{dn}(bx,k)}\Big)$.
It is easily seen that this potential respects $\cal PT$ -symmetry. The obvious
question is if the $\cal PT$-symmetry is spontaneously unbroken or not.
Unfortunately, since the periodic two-soliton solution is not obtained through
the IST method, one does not have the information about the corresponding
 eigenvalues. However, there are two reasons because of which we believe that
in this case too the $\cal PT$-symmetry is not spontaneously broken. Firstly, in the
limit when the periodic two-soliton solution goes over to the two-soliton
solution (\ref{f6b}), we know that the $\cal PT$-symmetry is not spontaneously
broken. Secondly, we have numerically calculated the low lying energy
eigenvalues and they turn out to be real. It would be nice if one can
rigorously show that even in this case the $\cal PT$-symmetry is not spontaneously
broken.

\begin{figure}[H]
\centering\includegraphics[width=0.4\linewidth]{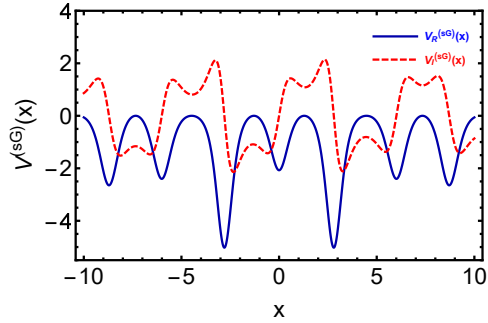}\\
\caption{Real part (solid blue line) and imaginary part (red-dashed line) of the periodic two-kink potential (\ref{f7b}) for $c=0.9$, $k=0.9$, $m=0.8$.}
\end{figure}
In Fig.~4, we display the real part (blue line) and imaginary part (red-dashed line) of periodic two-kink potential.
%This package mainly contains EigenNDSolve which employs a spectral expansion in terms of Chebyshev polynomials.

\subsubsection{ \bf Periodic breather potential }

In what follows, we discuss the construction of the periodic breather
potential following from the periodic breather solution of the sG equation.
The periodic breather solution of the sG equation is given by \cite{Fu}
\bes\bea\label{f7e}
u(x,t) = 4\tan^{-1}[\mbox{cd}(ax+\delta_{5}(t), k)~\mbox{dn}(cx+\delta_{6}(t), m)]\,.
\eea
This solution (\ref{f7e}) reduces to the breather solution (\ref{f6c}) in the
limit  $m\rightarrow1$ and $k\rightarrow0$. In the above solution, $a=c$;
$b=-\frac{1}{2(2-m^{2})a}$; $d=\pm\frac{1}{2(2-m^{2})a}$ and we consider
suitable initial time $t_{0}$ so as to make $\delta_{5}(t_{0})=bt_{0}+s_{0}$
and $\delta_{6}(t_{0})=dt_{0}+r_{0}$ to be zero. Then the solution (\ref{f7e})
can be rewritten as
\bea\label{f7f}
u(x) = 4\tan^{-1}[\mbox{cd}(ax, k)~\mbox{dn}(cx, m)].
\eea\ees
The complex potential resulting for the above solution (\ref{f7f}) is
\bea\label{f7g}
V^{(sG)}(x)=-\frac{1}{4}\Bigg[\Bigg(\frac{4 a (k-1) \mbox{dn}(cx, m)~\zeta_{3}-c~m~\mbox{cd}(ax, k)~\zeta_{4}}{1+\mbox{cd}^{2}(ax, k)~\mbox{dn}^{2}(cx, m)}\Bigg)^{2}+\nonumber\\
2i~\frac{\mbox{d}}{\mbox{dx}}\Bigg(\frac{4 a (k-1) \mbox{dn}(cx, m)~\zeta_{3}-c~m~\mbox{cd}(ax, k)~\zeta_{4}}{1+\mbox{cd}^{2}(ax, k)~\mbox{dn}^{2}(cx, m)}\Bigg)\Bigg].\nonumber\\
\eea
Here the parameters $\zeta_{3}$ and $\zeta_{4}$ can be expressed as $\mbox{nd}(ax, k)~\mbox{sd}(ax, k)$ and $\mbox{cn}(cx,m)~\mbox{sn}(cx,m)$. It is clear that this complex periodic breather-type potential does not have $\cal PT$-symmetry. This non-$\cal PT$ -symmetric potential varies its profile
as the parameters  $a$ and $c$ are changed.
\begin{figure}[H]
\centering\includegraphics[width=0.4\linewidth]{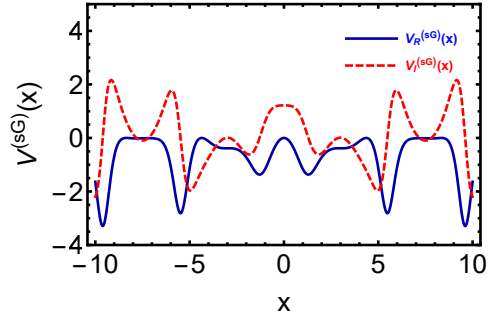}\\
\caption{Real part (solid blue line) and imaginary part (red-dashed line) of the periodic breather potential (\ref{f7g}) for $a=c=0.9$, $b=0.4$, $m=0.8$.}
\end{figure}
In Fig.~5, we display the real part (blue line) and imaginary part (red-dashed line) of periodic breather potential. This clearly shows the non-$\cal PT$ symmetric nature of potential.

 \section{Construction of  complex potentials from the solution of  modified KdV equation}

After the detailed discussion about the various complex potentials following from
the sG equation, we extend our approach to the integrable mKdV
equation (\ref{f8}).  As mentioned in the introduction the construction of the $\cal PT$ -symmetric
potentials for the mKdv equation has already been discussed by Wadati  in his
pioneering work \cite{wadati}. However, it is limited to the case of one and
two soliton potentials. It is of interest to construct the complex potentials
from the solution of mKdV equation by considering various other interesting non-solitonic
solutions like bion, elliptic bion and other periodic solutions \cite{drazin}.
Due to the physical
importance of the above mentioned solutions, we extend the elegant theory
developed by Wadati to construct  potentials corresponding to other
interesting solutions of the mKdV equation (\ref{f8}).

Then by following the procedure outlined in section 2.1, we obtain the
following complex potential for the linear Schr\"odinger equation by using the solution of the mKdV equation
\bea\label{f9}
 V^{(mKdV)}&=&V^{(mKdV)}_{R}+i V^{(mKdV)}_{I}\equiv-q^{2}-i q_{x}.
\eea
Note that as in the sG case, this complex potential is generated by the real
solution $q(x,t)$ of the mKdV equation.
In what follows, we deal with special elliptic bion solution, bion solution, dnoidal, cnoidal and superposed elliptic solutions of the mKdV system.

\subsection{ \bf  Bion potential}
We now consider the following interesting bion solution of the mKdV
equation \cite{drazin}.
\bes\bea\label{f12b}
q(x,t)=-2\frac{\mbox{d}}{\mbox{dx}}\Big[\tan^{-1}(a~\sin(\alpha x
+\rho_{4}(t))~\mbox{sech}(\beta x+\rho_{5}(t)))\Big]\,.
\eea
Here $\rho_{4}(t)$ and
$\rho_{5}(t)$ are given as $\upsilon t+a_{0}$ and $\varsigma t+ b_{0}$
respectively. The constraint conditions are $a=-\frac{\beta}{\alpha}$, $\upsilon=(\alpha^{2}-3\beta^{2})\alpha$ and $\varsigma=(3\alpha^{2}-\beta^{2})\beta$. As before, by choosing the initial time $t_{0}$ such that  $\rho_{4}(t_{0})=\upsilon t_{0}+a_{0}$
and $\rho_{5}(t_{0})=\varsigma t_{0}+ b_{0}$ vanish, one can obtain the following
stationary form
\bea\label{f12c}
q(x)=-2\frac{d}{d x}\Big[\tan^{-1}(a~\sin(\alpha x)~\mbox{sech}(\beta x))\Big]\,.
\eea\ees
This bion solution results in the following complex potential
\bea\label{f12ad}
V^{(mKdv)}(x)=\Bigg[-\Bigg(\frac{2a~\mbox{sech}(\beta x)(\alpha \cos(\alpha x)-\beta \sin(\alpha x)\tanh(\beta x))}{(1+a^2\mbox{sech}^2(x\beta) \sin^2(x\alpha))}\Bigg)^2\nonumber\\
-i \frac{\mbox{d}}{\mbox{dx}}\Bigg(\frac{2a\mbox{sech}(\beta x)(\alpha \cos(\alpha x)-\beta \sin(\alpha x)\tanh(\beta x))}{(1+a^2\mbox{sech}^2(x\beta) \sin^2(x\alpha))}\Bigg)\Bigg].
\eea
The above bion potential also possesses $\cal PT$ symmetry. Since it is not derived by
the IST method, we do not know for sure its energy eigenvalues and hence
it is not clear if the $\cal PT$-symmetry remains unbroken or not. We have numerically evaluated
low lying eigenvalues and they are real thereby suggesting that in this case too
the $\cal PT$-symmetry is not spontaneously broken. It would be nice if one can
rigorously prove it.
\begin{figure}[H]
\centering\includegraphics[width=0.4\linewidth]{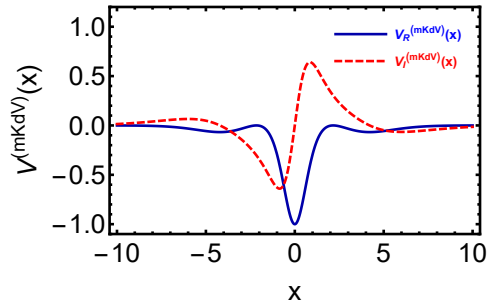}
\caption{Real part (solid blue line) and imaginary part (red-dashed line) of bion potential for $\alpha=0.5$ and $\beta=0.3$.}
\end{figure}
A plot of the potential is given in Fig. 6 from where too it is clear that
while the real part of the potential is symmetric, the imaginary part is
antisymmetric.

\subsection{ \bf  Elliptic bion potential}

We consider the following elliptic bion solution of the mKdV equation \cite{drazin}.
\bes\bea\label{f12}
q(x,t)=-2\frac{\mbox{d}}{\mbox{dx}}\Big[\tan^{-1}(a~\mbox{sn}(\alpha x+\rho_{4}(t),k)~\mbox{dn}(\beta x+{\rho_{5}(t)},m))\Big].
\eea
The parameters of solution (\ref{f12}) satisfy the relations, $a=-\frac{\beta}{\alpha}$; $\frac{\beta}{\alpha}=\left(\frac{k}{(1-m)}\right)^{\frac{1}{4}}$. Here the parameters $\rho_{4,5}(t)$ are given as $\upsilon t+a_{0}$ and $\varsigma t+ b_{0}$
respectively, in which the $\upsilon=(\alpha^{2}(1+k)-3\beta^{2}(2-m))\alpha$ and $\varsigma=(3\alpha^{2}(1+k)-\beta^{2}(2-m))\beta$.  By considering initial time $t_{0}$ such that  $\rho_{4}(t_{0})
=\upsilon t_{0}+a_{0}$ and $\rho_{5}(t_{0})=\varsigma t_{0}+ b_{0}$ vanish, we
obtain the stationary version of (\ref{f12}) which reads as
\bea
q(x)=-2\frac{\mbox{d}}{\mbox{dx}}\Big[\tan^{-1}(a~\mbox{sn}(\alpha x,k)~
\mbox{dn}(\beta x,m))\Big]\,.
\eea\ees
This elliptic bion solution results in the following complex potential
\bea\label{f12a}
V^{(mKdV)}(x)=\Bigg[-\Bigg(\frac{2a~(m~\beta~\gamma_{2}~\mbox{sn}(~\beta x,m)-\alpha~\gamma_{1}~\mbox{dn}(\beta x,m))}{(1+a^2\gamma^2)}\Bigg)^2\nonumber\\
-i \frac{d}{dx}\Bigg(\frac{2a~(m~\beta~\gamma_{2}~\mbox{sn}(~\beta x,m)-\alpha~\gamma_{1}~\mbox{dn}(\beta x,m))}{(1+a^2\gamma^2)}\Bigg)\Bigg].
\eea
Here the expressions for $\gamma$, $\gamma_{1}$ and $\gamma_{2}$~are $\mbox{dn}(\beta x,m)~\mbox{sn}(\alpha x,k)$,  $\mbox{cn}~(\alpha x,k)~\mbox{dn}(\alpha x,k)$ and
$\mbox{cn}(\beta x,m)~\mbox{sn}(\alpha x,k)$ respectively. It is easy to see that this
elliptic bion potential is $\cal PT$ -symmetric. Since this solution is not derived by the
IST method.
We have no clue about the nature of its energy eigenvalues. Hence
it is not clear if the $\cal PT$-symmetry remains unbroken or not. We have numerically evaluated
few low lying eigenvalues which turn out to be real thereby suggesting that for this case also
the $\cal PT$-symmetry is not spontaneously broken. A rigorous proof of this could be an interesting study.
\begin{figure}[H]
\centering\includegraphics[width=0.38\linewidth]{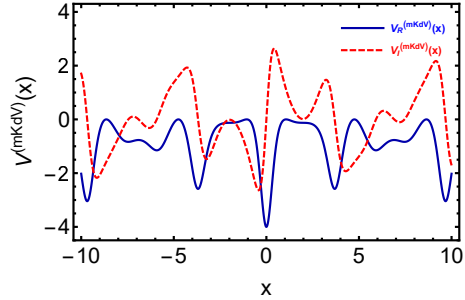}
\caption{Real part (solid blue line) and imaginary part (red-dashed line) of elliptic bion potential (\ref{f12a}) for $\alpha=1$; $m=0.8$ and $k=0.2$.}
\end{figure}
This elliptic bion potential is shown in Fig.~7 and it alters its profile as the
parameters  $\alpha$ and $\beta$ are varied as shown in Fig. 7.

\subsection{\bf Periodic potentials}

The above construction procedure of complex potential from the solution of the mKdV system
(\ref{f8}) can be straightforwardly extended to the other periodic
potentials like  dnoidal, cnoidal and superposed elliptic solutions of the
mKdV equation. We tabulate below these remaining solutions and the corresponding
potentials for brevity in Table 1.

 \renewcommand{\floatpagefraction}{0.1}
\renewcommand{\arraystretch}{0.1}% Tighter
\begin{table}[H]
\centering
\caption {Novel types of complex potentials for the mKdV equation}
\begin{tabular}{|l| l| l| l |l|}
\hline
Type of&$q(x,t)$& $q(x)$&$V^{(mKdV)}(x)$&Constraints\\
solutions & & & &\\
  \hline
dnoidal&$\mbox{A}~\mbox{dn}(\theta,m)$&$\mbox{A}~\mbox{dn}(\beta x,m).$&$\Big[-\mbox{A}^{2}~\mbox{dn}^{2}(\beta x,m)$&$A^{2}=\beta^{2}$\\
solution& & &$+i \mbox{A}~\mbox{m}~\beta \mbox{cn}(\beta x,m)$&$v=(2-m)\beta^{2}$\\
~&&&$\mbox{sn}(\beta x,m)\Big]$&\\
\hline
cnoidal&$\mbox{A}~\sqrt{m}~\mbox{cn}(\theta,m)$&$\mbox{A}~\sqrt{m}~\mbox{cn}(\beta x,m).$&$\Big[-\mbox{A}^{2}\mbox{m}~\mbox{cn}^{2}(\beta(x),m)$&$A^{2}=\beta^{2}$\\
solution&& &$+ i \mbox{A}~\sqrt{m}~\beta \mbox{dn}(\beta(x),m)$&$v=(2m-1)\beta^{2}$\\
&&&$\mbox{sn}(\beta(x),m)\Big]$&\\
\hline
Superposed&$\frac{A}{2}\Big(\mbox{dn}(\theta,m)\pm$&$\frac{A}{2}\Big(\mbox{dn}(\beta x,m)\pm$&$-\Bigg[\frac{A^{2}}{4}\Big(\mbox{dn}(\beta(x),m)$&\\
&&&$\pm\sqrt{m}\mbox{cn}(\beta(x),m)\Big)^{2}$&$~A^{2}=\beta^{2}$\\
solution&$\sqrt{m}\mbox{cn}(\theta,m)\Big)$&$\sqrt{m}\mbox{cn}(\beta x,m)\Big)$ &$\frac{iA}{2}\frac{\mbox{d}}{\mbox{dx}}\Big(\mbox{dn}(\beta(x),m)$&$v=\frac{(1+m)\beta^{2}}{2}$\\
&&&$\pm\sqrt{m}\mbox{cn}(\beta(x),m)\Big)\Bigg]$&\\
~&&&&\\
\hline
\end{tabular}
\end{table}
In Table 1, $\theta=\beta(x+\rho_{5}(t))$, where $\rho_{5}(t)$ can be
expressed as $(-v t+ x_{0})$ ($x_{0}$ is a constant). In the second column of
the Table 1, we present the time dependent elliptic solutions and we convert
those solutions as time independent solutions (stationary solutions) in the
third column by choosing the initial time $t_{0}$ appropriately such that
$\rho_{4}(t_{0})=-v t_{0}+ x_{0}=0$.

All the three elliptic potentials presented in Table 1 are $\cal PT$ -symmetric potentials. The obvious
question is if the $\cal PT$-symmetry is spontaneously unbroken or not.
Unfortunately, since these periodic soliton solutions are not obtained through
the IST method, the nature of the corresponding  eigenvalues can not be predicted. However, for two reasons we believe for these cases  the $\cal PT$-symmetry not to be spontaneously broken. Firstly, in the limit $m =1$ when these periodic soliton solutions goes over to the one soliton
solution and  from the IST procedure we know that the eigenvalue is real and hence
$\cal PT$-symmetry is not spontaneously broken. Secondly, we have numerically
calculated the low lying energy eigenvalues for these three $\cal PT$-invariant
potentials and they turn out to be real. It would be nice if one can
rigorously show that even in these cases the $\cal PT$-symmetry is not spontaneously
broken.

\begin{figure}[H]
\centering\includegraphics[width=0.32\linewidth]{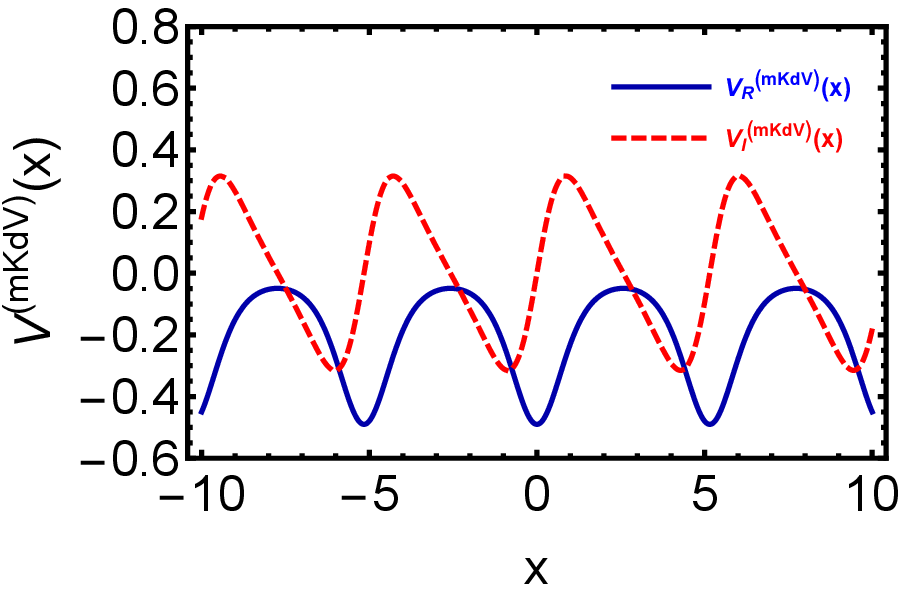}~\includegraphics[width=0.32\linewidth]{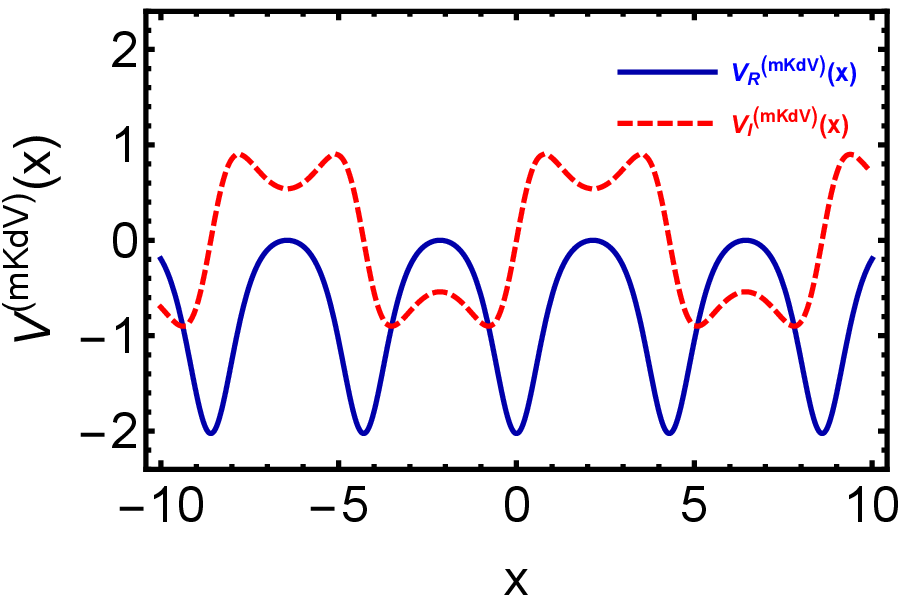}
\includegraphics[width=0.32\linewidth]{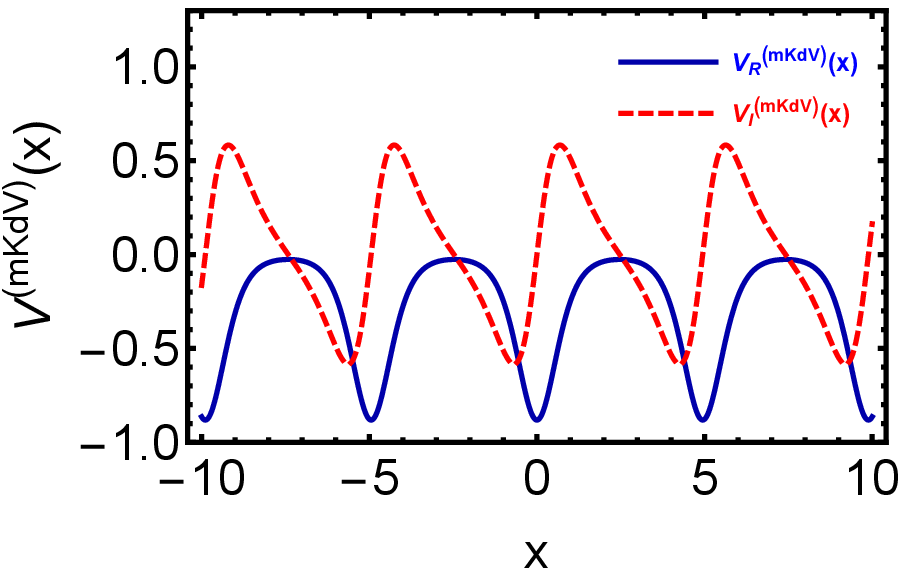}
\caption{Real part (solid blue line) and imaginary part (red-dashed line) of the dnoidal potential (Left panel) for $A=0.7$, $\beta=1$ and $m=0.9$; cnoidal potential (middle-panel) for $A=1.5$, $\beta=1.2$ and $m=0.9$ and superposed potential (right-panel) for $A=1.1$, $\beta=1.5$ and $m=0.5$ respectively.}
\end{figure}
In Fig.~8 we have depicted the real and imaginary parts of the dnoidal
(left-panel), cnoidal (middle-panel) and superposed potential (right-panel). All these potentials posses robust shape even
if we change the corresponding solution parameters.

\section{Construction of complex potentials from the solution of  mKdV-sG equation}

Now we extend our systematic construction of complex
potentials from the solution of combined mKdV-sG system which is receiving attention
recently \cite{leblond1}. As mentioned in the introduction, this system
describes the propagation of few-optical-cycle pulses in a self-focusing Kerr
medium which is of paramount of interest. The complete integrability of this
physically interesting system has been proved by means of the IST method
\cite{leblond1}. The mKdV-sG system is casted as
\bea\label{fms1}
u_{xt}+\Upsilon\left(\frac{3}{2}(u_{x})^{2}u_{xx}+u_{xxxx}\right)
-\Upsilon_{1} \sin(u)=0\,,
\eea
where the field $u$ is a function of $x$ and $t$. The constants $\Upsilon$ and
$\Upsilon_{1}$ account for the dispersion and nonlinearity properties of the
medium and are chosen to be one without loss of generality. The Lax pair for the
system (\ref{fms1}) is well known \cite{leblond1}. The spatial evolution
equations for the mKdV-sG system and the sG system are identical albeit their
time evolution differs. Hence the complex potential for the linear Schr\"odinger equation using solution of this mKdV-sG equation can be obtained following the procedure for the sG system given in section 2.
In particular
\bes\bea\label{fms3}
V^{(mKdv-sG)}(x)&=&V^{(mKdv-sG)}_{R}(x)+i V^{(mKdV-sG)}_{I}(x),
\eea
where the constituent real and imaginary parts of the  $V^{(mKdV-sG)}(x)$ are
\bea
 V^{(mKdV-sG)}_{R}(x)&=&-\frac{u_{x}^{2}}{4},\\
 V^{(mKdV-sG)}_{I}(x)&=&-\frac{u_{xx}}{2}.
\eea\ees
Note that here the quantity $u$ corresponds to the solution of (\ref{fms1}).
Following the sG example, we can similarly correlate the potential of the
graphene model with the solution of the mKdV-sG system. Not surprisingly,
the connection between the potential of the graphene model and the solution of
the mKdV-sG system looks similar to the sG case and is given by:
\bea\label{fms4}
V(x)=\frac{u_{x}}{2}\,,
\eea
\noindent where $u$ is the solution of the mKdV-sG system (\ref{fms1}).
Hence the form of the potential will be entirely different from that of the sG
case. This shows the existence of a host of complex exactly solvable potentials
 for the graphene model.\\

\noindent Like the sG case, for the integrable mKdV-sG equation too people have obtained
N-kink soliton solution through the IST method. It turns out that in
spite of having different mathematical form  from the sG
system mKdV-sG system still admits the same stationary form for the one kink soliton as
the sG  case which is given in Eq. (\ref{f6a}). So, here we skip the discussion about
the one-kink soliton of the mKdv-sG system. However, the two- kink soliton of
the mKdV-sG system has distinct form that of the sG system.

\subsubsection{ \bf Two kink potential }
In this subsection, we focus on the construction of the two kink
$\cal PT$ -symmetric potential from the two kink solution of the mKdV-sG
equation. This two kink solution of the mKdV-sG equation (\ref{fms1}) reads as
\bes\bea
u(x,t) = 4\tan^{-1}\left[ \frac{a_{11} \cosh((\eta_{2}-\eta_{1})x+\Lambda_{1}(t))}{\sinh((\eta_{1}+\eta_{2})x+\Lambda_{2}(t))}\right],
\label{fms4a}\eea
where $a_{11}=\sqrt{\frac{(\eta_{1}+\eta_{2})^{2}(c_{10}\eta_{1}+c_{20}\eta_{2})}{c_{10}c_{20}(\eta_{1}-\eta_{2})^{2}}}$ with $\eta_{1,2}$ being the  solion parameters. The expressions for $\Lambda_{1}(t)$ and
$\Lambda_{2}(t)$ are given by $(A_{20}-A_{10})t+s_{1}$ and $(A_{10}+A_{20})
+r_{1}$, respectively in which $s_{1}$ and $r_{1}$ are expressed as
$\frac{1}{2}\ln(\frac{c_{10}}{2\eta_{1}}-\frac{c_{20}}{2\eta_{2}})$ and
$\frac{1}{2}\ln(\frac{c_{10}c_{20}(\eta_{1}-\eta_{2})^{2}}{4\eta_{1}
\eta_{2}(\eta_{1}+\eta_{2})^{2}})$, respectively. Here $c_{10}$ and $c_{20}$
are real arbitrary parameters. One can easily reduce the following stationary
solution from the solution (\ref{fms4a}) by considering the initial time
$t_{0}$ as $\Lambda_{1}(t_{0})=(A_{20}-A_{10})t_{0}+s_{1}\equiv0$ and
$\Lambda_{2}(t_{0})=(A_{10}+A_{20})t_{0}+r_{1}\equiv0$:
\bea\label{fms5}
u(x)=4\tan^{-1}\left[ \frac{a_{11}\cosh((\eta_{2}-\eta_{1})x)}{\sinh((\eta_{1}+\eta_{2})x)}\right].\eea\ees
Using solution (\ref{fms5}), we obtain the following complex potential
\bes\bea\label{fms6}
V^{(mKdV-SG)}(x)=V^{(mKdV-sG)}_{R}(x)+i V^{(mKdV-sG)}_{I}(x),
\eea
where the real and imaginary parts of the two kink potential from the two kink solution of mKdV-sG
equation are expressed as
\bea\label{fms7}
\fl V^{(mKdV-sG)}_{R}(x)=-\frac{1}{4}\Bigg[\frac{a_{11}\mbox{csch}(b_{12}x)\Big(b_{11}~\sinh(b_{11}x)
-b_{12}~\cosh(b_{11}x)\coth(b_{12}x)\Big)}{1+\Big(a_{11}\frac{\cosh(b_{11}x)}{\sinh(b_{12}x)}\Big)^{2}}\Bigg]^{2},\nonumber\\
\eea
\bea\label{fms7d}
\fl V^{(mKdV-sG)}_{I}(x)= -\frac{1}{2}\frac{\mbox{d}}{\mbox{dx}}\Bigg[\frac{a_{11}\mbox{csch}(b_{12}x)\Big(b_{11}~\sinh(b_{11}x)
-b_{12}~\cosh(b_{11}x)\coth(b_{12}x)\Big)}{1+\Big(a_{11}\frac{\cosh(b_{11}x)}{\sinh(b_{12}x)}\Big)^{2}}\Bigg].\nonumber\\
\eea\ees
Here $b_{11}=(\eta_{2}-\eta_{1})$,\,\,$b_{12}=(\eta_{1}+\eta_{2})$. It is easy to check that this two-kink potential (\ref{fms6}) associated with mKdV-sG
equation respects $\cal PT$ -symmetry and in fact the $\cal PT$-symmetry is not
spontaneously broken. This is because, in the case of the two kink
solution of the mKdV-sG system, it is well known that
the spectral parameters ($\zeta_{1}$, $\zeta_{2}$) associated with the kink solution (\ref{fms4a}) are purely imaginary ($i \eta_{1}$, $i \eta_{2}$) thereby resulting in real energy eigenvalues, $E_{1}=-\eta_{1}^{2}$ and $E_{2}=-\eta_{2}^{2}$.
\begin{figure}[H]
\centering\includegraphics[width=0.4\linewidth]{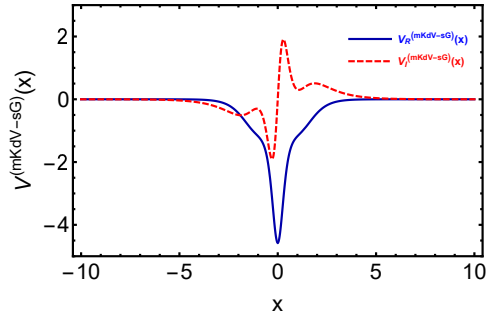}
\caption{Real part (solid blue line) and Imaginary part (red-dashed line) of two kink potential from the two kink solution of the mKdV-sG equation for $c_{10}=0.5$, $c_{20}=1$, $\eta_{1}=0.5$ and $\eta_{2}=2.5$.}
\end{figure}
In Fig.~9 we plot the real and the imaginary parts of the two kink potential associated with mKdV-sG equation.

\section{Construction of  complex potentials from the solutions of Gardner equation}

Finally, we shift our attention to another important soliton bearing
integrable nonlinear model, i.e.  the Gardner equation (GE). This equation is an
extended version of the KdV equation with cubic nonlinearity.
The integrability nature of the Gardner equation has been  studied by using
the IST method \cite{wadati1}. The $\cal  PT$ -symmetric potential from the solution of the
Gardner equation corresponding to the  one-soliton solution has already been
obtained in Ref. \cite{wadati}. However, these are other interesting solutions of the
Gardner equation such as two-soliton solution, breather and elliptic
solutions and it is worth constructing complex potentials corresponding to
some of these solutions. This is what we do in this section.

We consider the following dimensionless form of the Gardner equation \cite{wazwaz}:
\bea\label{f15}
u_{t}+6 \alpha_{1} u u_{x}+6\beta_{1} u^{2}u_{x}+u_{xxx}=0.
\eea
Here $\alpha_{1}$ and $\beta_{1}$ are the nonlinearity coefficients. In order
to construct the complex potentials of the Gardner equation, we consider its
Lax representation. The spatial and temporal evolution equations of the
Gardner equations are
\bes \label{f15a}\bea\label{f15b}
v_{1x}+ i \zeta v_{1}&=&q v_{2}, \\\label{f15c}
v_{2x}-i \zeta v_{2}&=& r v_{1}
\eea
and
\bea\label{f15d}
v_{1t}=(- 4\eta^{3}+2\eta q r+r q_{x}-q r_{x}) v_{1}+ (-4\eta^{2}q-2\eta q_{x}+2q^{2}r-q_{xx}) v_{2},\\
v_{2t}=(-4\eta^{2}r-2\eta r_{x}+2q r^{2}-r_{xx})v_{1}-(- 4\eta^{3}+2\eta q r+r q_{x}-q r_{x}) v_{2},
\eea\ees
where $q=u$ and $r=-(\alpha_{1}+\beta_{1} u)$. Following the procedure as
outlined in section 2.1, one can construct the complex potentials for the linear Schr\"odinger equation using the solution of the Gardner equation
\bes\bea\label{f16}
V^{(G)}&=&V^{(G)}_{R}+ i V^{(G)}_{I},
\eea
where the real and imaginary parts of the potentials are given by
\bea\label{f16a}
 V_{R}^{(G)}&=&-\alpha_{1} u-\beta_{1} u^{2},\\
 V_{I}^{(G)}&=&-\sqrt{\beta_{1}}u_{x}.
\eea\ees

\subsection{\bf Two soliton potential}

The two soliton solution of the Gardner equation has been derived using the
Hirota's bilinearization method. The form of the two soliton solution
of the Gardner equation is \cite{wazwaz}
\bes\bea\label{f17}
\fl u(x,t)=q_{0}\left[\frac{\cosh(k_{2}x -\rho_{7}(t))+\cosh(k_{1}x -\rho_{8}(t))}{\cosh(\kappa x -\rho_{9}(t))+\cosh(\kappa_{1}x -\rho_{10}(t))\cosh(\kappa_{1}x-\rho_{11}(t))}\right]-\frac{1}{2}.\nonumber\\
\eea
Here, $q_{0}=2\sqrt{k_{1}k_{2}a_{12}}$; $\rho_{7,8}(t)$, $\rho_{9}(t)$, $\rho_{10}(t)$ and $\rho_{11}(t)$ ~are
expressed as $(k_{2,1}^{3}-\frac{3}{2}k_{2,1})t+\tau_{2,1}$ and
$((k_{1}^{3}+ k_{2}^{3})-\frac{3}{2}(k_{1}+ k_{2}))t-\tau_{3} )$,
$\frac{1}{2}((k_{1}^{3}- k_{2}^{3})-\frac{3}{2}(k_{1}- k_{2}))t-\tau_{4} )$
and $\frac{1}{2}((k_{1}^{3}- k_{2}^{3})-\frac{3}{2}(k_{1}- k_{2}))t+\tau_{4} )$
 respectively, in which $\tau_{1}=\frac{(e_{1}+f)}{2}$, $\tau_{2}
=\frac{(e-f_{1})}{2}$, $\tau_{3}=c_{1}$   and $\tau_{4}=c$. The various quantities are
defined as  $e=\log(k_{1})$, $e_{1}=\log(k_{2})$, $f=\log(a_{12}k_{1})$,
$f_{1}=\log(a_{12}k_{2})$, $c=(1-a_{12})$, $c_{1}=\frac{a_{12}^{2}}{2}$,
$\kappa=(k_{1}+k_{2})$, $\kappa_{1}=(k_{1}-k_{2})$  and
$a_{12}=\frac{(k_{1}-k_{2})}{(k_{1}+k_{2})}$. Choosing the initial time
$t_{0}$ suitably so that  $\rho_{j}(t_{0})=0$, $j=7, 8, 9, 10, 11$ which
requires $(k_{2}^{3}-\frac{3}{2}k_{2})t_{0}+\tau_{2}=0$, $(k_{1}^{3}
-\frac{3}{2}k_{1})t_{0}+\tau_{1}=0$, $((k_{1}^{3}+ k_{2}^{3})-\frac{3}{2}(k_{1}
+ k_{2}))t_{0}-\tau_{3} )=0$, $\frac{1}{2}((k_{1}^{3}- k_{2}^{3})
-\frac{3}{2}(k_{1}- k_{2}))t_{0}-\tau_{4} )=0$ and $\frac{1}{2}((k_{1}^{3}
- k_{2}^{3})-\frac{3}{2}(k_{1}- k_{2}))t_{0}+\tau_{4} )=0$,  the solution
(\ref{f17}) becomes
\bea
u(x)=q_{0}\left[\frac{\cosh(k_{2}x)+\cosh(k_{1}x )}{\cosh(\kappa x)+\cosh^{2}(\kappa_{1}x)}\right]-\frac{1}{2}.
\eea\ees
One can easily construct the following complex two soliton potential by
using Eq. (37)
\bes\bea
V^{(G)}(x)&=&V^{(G)}_{R}(x)+ i V^{(G)}_{I}(x),
\eea
where the real and imaginary parts of the two soliton potential are:
\bea
V_{R}^{(G)}=-\alpha_{1} \left(q_{0}\left[\frac{\cosh(k_{2}x)+\cosh(k_{1}x )}{\cosh(\kappa x)+\cosh^{2}(\kappa_{1}x)}\right]-\frac{1}{2}
\right)\nonumber\\
-\beta_{1}\left(q_{0}\left[\frac{\cosh(k_{2}x)+\cosh(k_{1}x )}{\cosh(\kappa x)+\cosh^{2}(\kappa_{1}x)}\right]-\frac{1}{2}\right)^{2}
\eea
and
\bea
V_{I}^{(G)}=-\sqrt{\beta_{1}}\frac{\mbox{d}}{\mbox{dx}}\left(q_{0}\left[\frac{\cosh(k_{2}x)+\cosh(k_{1}x )}{\cosh(\kappa x)+\cosh^{2}(\kappa_{1}x)}\right]-\frac{1}{2}\right).
\eea\ees
This two soliton potential from the two soliton solution of the Gardner equation is  $\cal PT$-symmetric but it is not obvious if $\cal PT$-symmetry is spontaneously unbroken or not. Numerically calculated low lying eigenvalues are real. This suggests that indeed $\cal PT$-symmetry remains
unbroken in this case. A rigorous proof of this is of interest in future.
\begin{figure}[H]
\centering\includegraphics[width=0.4\linewidth]{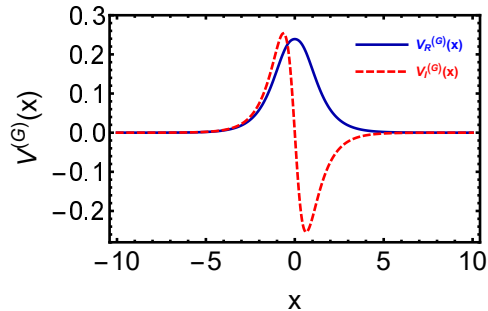}
\caption{Real part (solid blue line) and imaginary part (red-dashed line) of two soliton potential from the two soliton solution of the Gardner equation for $k_{1}=1.5$, $k_{2}=1.1$, $\alpha_{1}=1$ and $\beta_{1}=1$.}
\end{figure}
Fig.~10 shows the real and imaginary parts of the two soliton potential from the two soliton solution of
the Gardner equation. The shape of this two soliton potential can be altered by
changing the values of the solution parameters $k_{1}$ and $k_{2}$.

Finally, we extend the above construction procedure of complex potentials from the periodic solution of
the Gardner equation to special dnoidal, cnoidal and superposed elliptic
solutions of the Gardner equation, which have been obtained by one of us
\cite{avinash} with $\gamma=6\alpha_{1}$, $\alpha=6\beta_{1}$ and $\delta=1$.
For brevity, we present these solutions and the corresponding complex
$PT$-invariant potentials in  Table 2.

\renewcommand{\floatpagefraction}{0.7}
\renewcommand{\arraystretch}{0.}% Tighter
\begin{table}[H]
\centering
\caption {Novel type of complex potential for the Gardner equation}
\begin{tabular}{|l |l| l| l |l|}
\hline
\small{Type of}&\small{$q(x,t)$}&\small{$q(x)$}&\small{$V^{(G)}(x)$}&\small{Parametric}\\
\small{solutions.} & & && \small{constraints}\\
  \hline
\small{dnoidal}&\small{$A_{1}+B_{1}\times$}&\small{$A_{1}+B_{1}\times$}&\small{$\Big[\alpha_{1}(A_{1}+B_{1}~\mbox{dn}(p x,m))$}&\small{$A_{1}=-\frac{\alpha_{1}}{2\beta_{1}}$}\\
&\small{$\mbox{dn}(\vartheta,m)$}&\small{$~\mbox{dn}(p x,m).$}&\small{$-\beta_{1}(A_{1}+B_{1}~\mbox{dn}(p x,m))^{2}$}&\small{$B_{1}^{2}=\frac{p^{2}}{\beta_{1}}$}\\
\small{solution}& &  &\small{$-i B_{1}~\mbox{m}~\sqrt{\beta_{1}}~p~\mbox{cn}(p x,m)$}&\small{$v_{1}=$}\\
&&&~\small{$\mbox{sn}(p x,m)$\Big]}&\small{$(2-m)p^{2}$}\\
&&&&\small{$-\frac{3\alpha_{1}^{2}}{2\beta_{1}}$}\\
\hline
\small{cnoidal}&\small{$A_{1}+B_{1}\times$}&\small{$A_{1}+B_{1}\times$}&\small{$\Big[\alpha_{1}(A_{1}+B_{1}\sqrt{m}~\mbox{cn}(p x,m))$}&\small{$A_{1}=-\frac{\alpha_{1}}{2\beta_{1}}$}\\
&~\small{$\sqrt{m}~\mbox{cn}(\vartheta,m)$}&\small{$\sqrt{m}~\mbox{cn}(p x,m).$}&\small{$-\beta_{1}(A_{1}+B_{1}\sqrt{m}~\mbox{cn}(px,m))^{2}$}&\small{$B_{1}^{2}=\frac{p^{2}}{\beta_{1}}$}\\
\small{solution}&& &\small{$- i~B_{1}~\sqrt{m\beta_{1}}~p~\mbox{dn}(px,m)$}&\small{$v_{1}=$}\\
&&&~\small{$\mbox{sn}(px,m)$\Big]}&\small{$(2m-1)p^{2}$}\\
&&&&\small{$-\frac{3\alpha_{1}^{2}}{2\beta_{1}}$}\\
\hline
\small{Superposed}&\small{$A_{1}+\frac{B_{1}}{2}\times$}&\small{$A_{1}+\frac{B_{1}}{2}\times$}&\small{$\Bigg[\frac{(\alpha_{1}}{2}~\Big(2A_{1}+B_{1}~\mbox{dn}(px,m)$}&\small{$A_{1}=-\frac{\alpha_{1}}{2\beta_{1}}$ }\\
&\small{$\mbox{dn}(\vartheta,m)+$}&\small{$~\mbox{dn}(p x,m)+$}&\small{$+D~\mbox{cn}(px,m)\Big)-\frac{\beta_{1}}{4}\Big(2A_{1}$}&\small{$B_{1}^{2}=\frac{p^{2}}{\beta_{1}}$} \\
\small{solution}&\small{$\frac{D}{2}\sqrt{m}~\mbox{cn}(\vartheta,m)$}&\small{$\frac{D}{2}\sqrt{m}~\mbox{cn}(p x,m)$} &\small{$+B_{1} ~\mbox{dn}(px,m)$}&\small{$D=\pm B_{1}$}\\
&&&\small{$+D~\mbox{cn}(px,m)\Big)^{2}$}&\small{$v_{1}=$}\\
&&&\small{$-i\frac{\sqrt{\beta_{1}}~p~\mbox{sn}(p x,m)}{2}$}&\small{$\frac{(1+m)p^{2}}{2}$}\\
&&&\small{$\Big(B_{1}~\mbox{dn}(px,m)$}&\small{$-\frac{3\alpha_{1}^{2}}{2\beta_{1}}$}\\
&&&\small{$+D~\mbox{cn}(px,m)\Big)\Bigg]$}&\\
\hline
\end{tabular}
\end{table}
In Table 2, $\vartheta=p(x+\delta_{0}(t))$, where $\delta_{0}(t)$ is given by
$(-v_{1} t+ x_{0})$ ($x_{0}$ is a constant). We exclude the time dependent
part in the third column by choosing the initial time $t_{0}$ such that
$\delta_{0}(t_{0})=-v_{1} t_{0}+ x_{0}=0$.

It is easily checked that all three potentials presented in Table 2 satisfy
$\cal PT$-symmetry.
\begin{figure}[H]
\centering\includegraphics[width=0.32\linewidth]{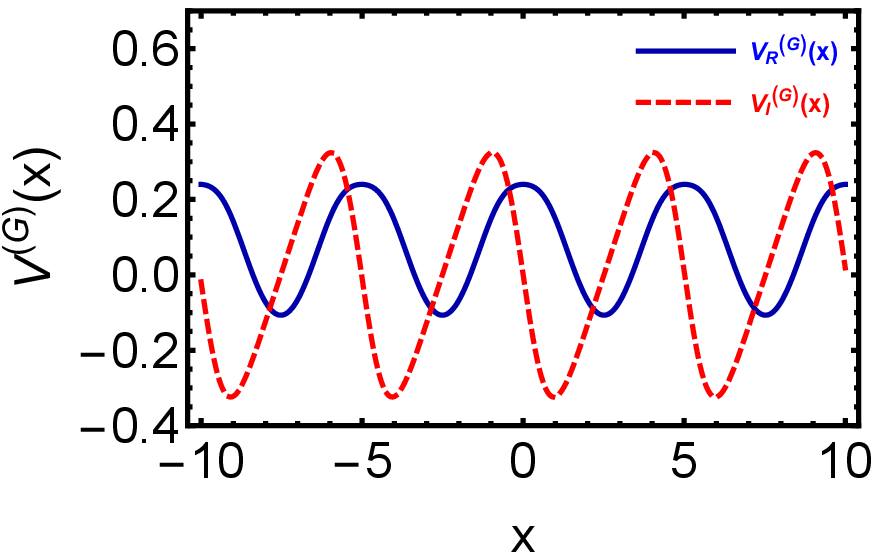}\includegraphics[width=0.32\linewidth]{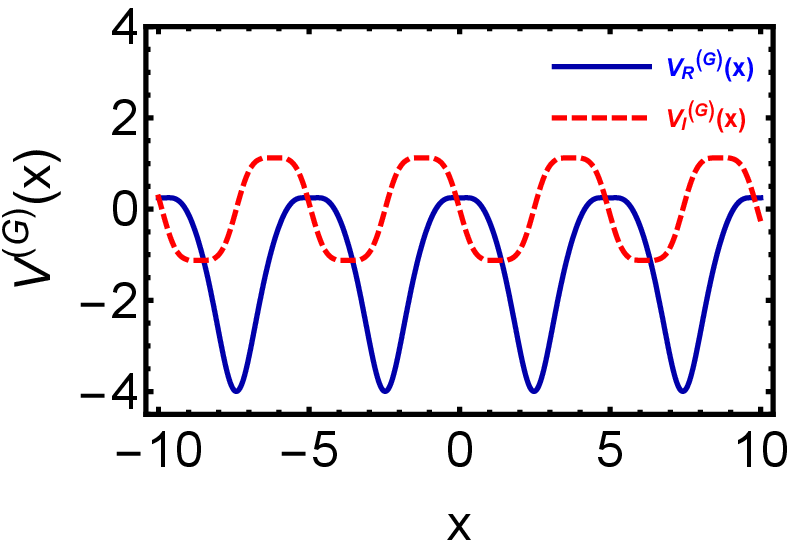}
\centering\includegraphics[width=0.32\linewidth]{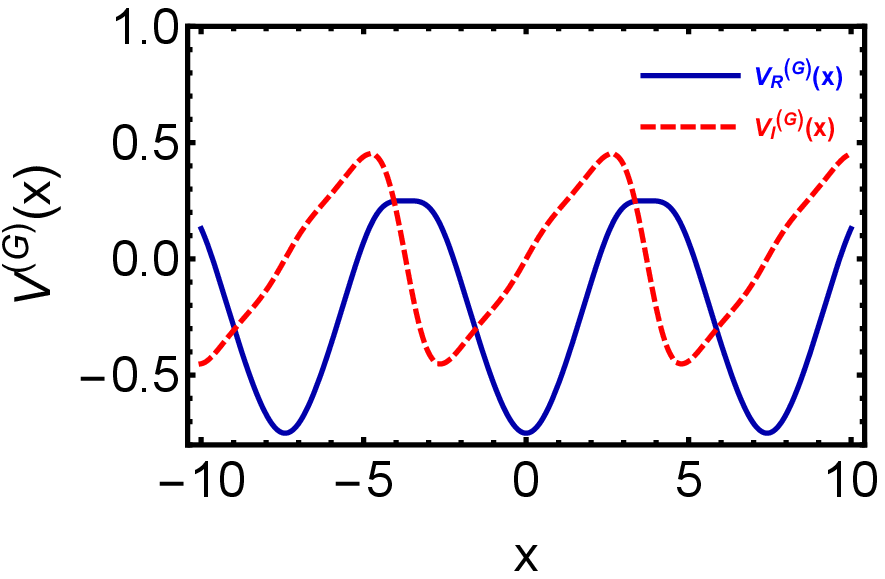}\\
\caption{Real part (solid blue line) and imaginary part (red-dashed line) of the elliptic complex potential of the dnoidal (left-panel) potential for $p=0.9$ and $m=0.8$; cnoidal (middle-pannel) potential for $p=1.5$ and $m=0.5$ and superposed (right-panel) potential for $p=1$ and $m=0.5$ respectively. In all the figures $\alpha_{1}=\beta_{1}=1$.}
\end{figure}
In Fig.~11, we depict the real and the imaginary parts of the dnoidal, cnoidal
and superposed potentials from the corresponding solution of the Gardner equation, respectively.
All these potentials have robust shape even if we alter the corresponding solution parameters.

\section{Conclusion}
In this paper, we have constructed a large number of complex potentials
including $\cal PT$ -symmetric potentials of the stationary Schr\"odinger equation from the various interesting soliton and periodic solution of the four fundamental and
ubiquitous real NLEEs, namely sG, mKdV, mKdV-sG and Gardner equations, based
on their corresponding Lax representation. We have considered several
different solutions of those equations namely soliton and periodic soliton
solutions and systematically constructed distinct classes of so called Wadati
potentials. Remarkably, in majority of cases these potentials turned out to be
$\cal PT$-symmetric while in the case of the breather or the periodic breather
solutions the corresponding potentials are not $\cal PT$-invariant. We also
investigate the robust nature of shape of the complex potentials by altering
the corresponding solution parameters. One of the nontrivial issues that we
have partially addressed in this paper is whether $\cal PT$-symmetry is
spontaneously broken or not in those potentials which are $\cal PT$-invariant.
For those cases where the solutions are derived using the IST method one knows
for sure that for the solitonic solutions the energy eigenvalues are real and
hence $\cal PT$ symmetry is not spontaneously broken. On the other hand, from the
IST method one knows that the energy eigenvalues are complex conjugate pair in
the case of the complex potentials following from the breather solution. In any
case for such complex conjugate eigenvalues  the complex potential does not admit $\cal PT$-symmetry. However, we have also obtained potentials by using periodic soliton solutions of these
equations. In this case we have no guidance from the IST method. Therefore
we have numerically calculated few low lying energy eigenvalues in all these
cases and remarkably we found in each and every case of this paper that
whenever the complex potential is $\cal PT$-invariant, then the energy
eigenvalues turn out to be real and hence $\cal PT$-symmetry is not
spontaneously broken. On the otherhand, whenever the complex potential is not
$\cal PT$-invariant, we found that the corresponding low lying energy
eigenvalues are complex thereby suggesting that in all the $\cal PT$-invariant
potentials derived in this way, the $\cal PT$-symmetry is not spontaneously broken.
It would be nice if one can rigorously prove this in all such cases.

A salient feature of our above study is that we have constructed complex
potentials by starting from real NLEEs. To the best of our knowledge all the
identified $\cal PT$ -symmetric potentials are new except for the Scarf-II
potential. We have also shown the possible relevance of these
$\cal PT$-symmetric potentials (and hence solutions) from the solution of the sG and mKdV-sG
systems in the context of the graphene model. Finally, we hope that the
various $\cal PT$ and non-$\cal PT$ -symmetric potentials obtained  here will
shed some light on $\cal PT$ -symmetric quantum mechanics.

\section*{Acknowledgments}
The work of K.T is supported by the National Fellowship for Scheduled Caste Students, University Grants Commission (UGC) of India. The work of T.K is supported by Department of Science and Technology-Science and Engineering Research Board (DST-SERB), Government of India, in the form of a major research project (File No. EMR/2015/001408). A.K. acknowledges Indian National Science Academy (INSA) for the award of INSA senior professorship at Savitribai Phule Pune University.

 \qquad\qquad\qquad\qquad\qquad\qquad\qquad\qquad***
   \end{document}